\documentclass[twocolumn,showpacs,groupedaddress]{revtex4-1}

\usepackage{color,amsmath,amssymb,amsthm,times,graphics,graphicx,bm,bbm,dcolumn}
\usepackage{epsfig}
\usepackage{graphicx}

\begin{document}

\title{Efficient all-optical production of large $^6$Li quantum gases using D$_1$ gray-molasses cooling}

\author{A.\ Burchianti$^{1,2}$, G.\ Valtolina$^{1,2,3}$, J.\ A.\ Seman$^{1,2}$, E.\ Pace$^4$, M.\ De Pas$^{2}$, M.\ Inguscio$^{2,5}$,  M.\ Zaccanti$^{1,2}$ and G.\ Roati$^{1,2}$}

\affiliation{$^{1}$INO-CNR, via Nello Carrara 1, 50019 Sesto Fiorentino, Italy}
\affiliation{$^{2}$LENS and Universit\`{a} di Firenze, Via Nello Carrara 1, 50019 Sesto Fiorentino, Italy}
\affiliation{$^{3}$Scuola Normale Superiore, Piazza dei Cavalieri, 7, 56126 Pisa, Italy}
\affiliation{$^{4}$Department of Physics, MIT-Harvard Center for Ultracold Atoms, and Research Laboratory of Electronics, Massachusetts Institute of Technology, Cambridge, Massachusetts, 02139, USA}
\affiliation{$^{5}$INRIM, Strada delle Cacce 91, 10135 Torino, Italy}

\begin{abstract}


We use a gray molasses operating on the D$_1$ atomic transition to produce degenerate quantum gases of $^{6}$Li with a large number of atoms. This sub-Doppler cooling phase allows us to lower the initial temperature of 10$^9$ atoms from 500 to 40~$\mu$K in 2~ms. We observe that D$_1$ cooling remains effective into a high-intensity infrared dipole trap where two-state mixtures are evaporated to reach the degenerate regime. We produce molecular Bose-Einstein condensates of up to 5$\times$10$^{5}$ molecules and weakly-interacting degenerate Fermi gases of $7\times$10$^{5}$ atoms at $T/T_{F}<0.1$ with a typical experimental duty cycle of 11 seconds.
\end{abstract}

\maketitle

\begin{figure}
\resizebox{0.4\textwidth}{!}
{  \includegraphics{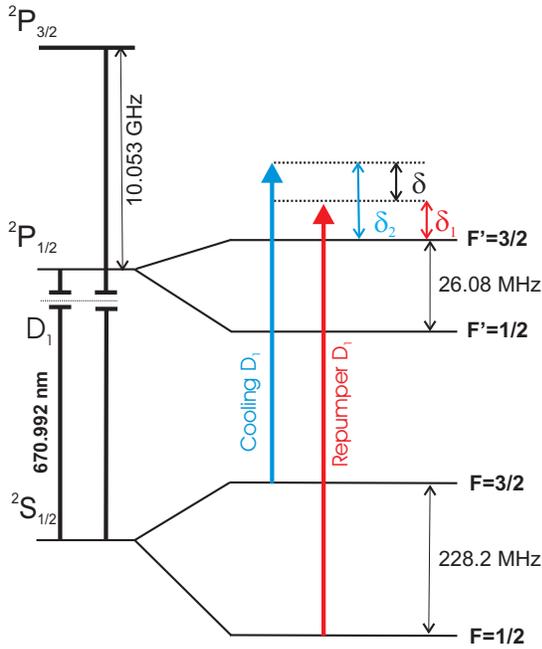}}
\caption{(Color online) Level scheme (not to scale) for $^{6}$Li showing the $D_{1}$ hyperfine structure and the transitions used for the gray-molasses. The laser detuning of the repumper and cooling lights are $\delta_{1}$ and $\delta_{2}$, while their relative detuning is $\delta$.
}
\label{fig:schemalivelli}     
\end{figure}

\begin{figure}
\includegraphics[width=7. cm,clip]{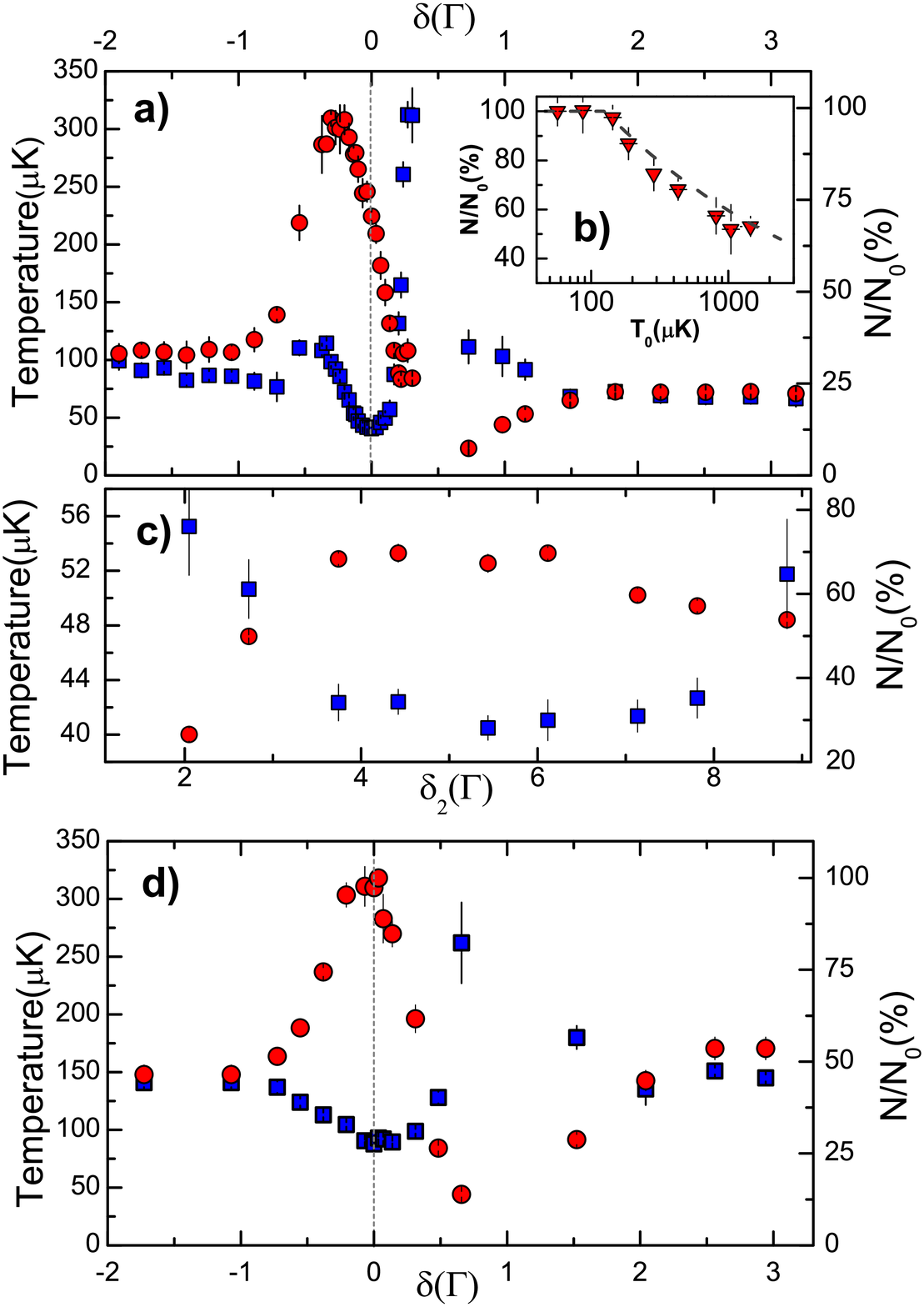}
\caption{(Color online) (a) Temperature (blue squares) and cooled fraction N/N$_0$ (red circles) after 2 ms of gray-molasses versus the relative detuning $\delta$ in units of $\Gamma$ with $\delta_{2}=5.4\Gamma$, $I_{cool}=2.7I_{S}$ and $I_{rep}=0.5I_{S}$. (b) N/N$_0$ versus the initial cloud temperature T$_0$ at $\delta$=0. Clouds with T$_0<$~200~$\mu$K are produced after 2 ms of gray-molasses with parameters away from the Raman condition. The dotted line is just a guide for the eye. (c) T (blue squares) and N/N$_0$ (red circles) versus the absolute detuning $\delta_2$, keeping $\delta$=0. (d) Same as (a), for a 300 $\mu$s molasses stage applied to atoms loaded into the optical trap at P=120~W (see text for details). In (a) and (d) the dotted line indicates the value $\delta$=0. In all the plots, the error bars are one standard deviation of five independent measurements.
}
\label{fig:Fano}       
\end{figure}

Ultracold atoms have emerged over the last decade as ideal quantum simulators of many-body phenomena, representing model systems to test quantum Hamiltonians \cite{Bloch_Rev}. In particular, the production of quantum gases of fermionic particles has opened new ways of studying  condensed matter problems with high controllability and unprecedented clarity \cite{varenna}.
The quest to develop new and efficient experimental schemes to produce large and highly degenerate fermionic samples is therefore a crucial challenge. To achieve this, all-optical schemes, as opposed to magnetic ones, are particularly appealing due to their higher flexibility \cite{Chapman}, allowing the trapping of any internal state also in the presence of magnetic fields. This is essential for efficient forced evaporation by exploiting magnetic Feshbach resonances \cite{Thomas, Grimm}. Implementing such a cooling strategy requires sufficiently dense and cold clouds to match the optical trap volume and depth, to ensure that a good fraction of the atoms are captured from the magneto-optical trap (MOT). The minimum theoretical temperature achievable in a MOT is typically restricted to the Doppler limit, $T_{D}=\hbar\Gamma/2k_{B}$, where $k_{B}$ is the Boltzmann constant, $\hbar$ is the reduced Planck constant and $\Gamma$ is the linewidth of the cooling transition. For most of alkali atoms, sub-Doppler cooling well below T$_D$ is generally achieved with Sisyphus cooling in optical molasses \cite{Dalibard-JOSA-1989}. For lithium isotopes, which are widely implemented in many experiments, the standard sub-Doppler mechanism is hindered by the unresolved splitting of the $2P_{3/2}$ level. Nonetheless, temperatures slightly below T$_D$=140~$\mu$K have been recently achieved for bosonic $^7$Li atoms, despite cooling only 45$\%$ of the initial sample \cite{lithium_Sys}.\\ 
Very cold MOTs in the tens of $\mu$K have been produced with both $^{6}$Li and $^{40}$K \cite{Hulet,Thywissen} by exploiting narrow transitions \cite{narrow} in the near UV region. However, this scheme requires special broadband optical components and eventually expensive laser sources.\\
In this paper, we present a simple and efficient way to prepare large fermionic $^6$Li quantum gases. Our scheme is based on the well-established D$_1$ (2S$_{1/2}\rightarrow$2P$_{1/2}$) gray-molasses cooling, 
so far successfully demonstrated only for bosonic $^7$Li \cite{PhysRevA.87.063411} and for potassium isotopes \cite{Chevy-EPL-12,PhysRevA.88.053407,Bourdel-EPLdraft-13}. We measure temperatures as low as 40~$\mu$K in the molasses without any significant reduction of the MOT atom number \cite{note0}. This allows an efficient transfer into an optical potential where we evaporate down to the degenerate regime. Remarkably, we continue to observe effective D$_1$ cooling in the high-intensity optical trap (peak-intensities of few MW/cm$^2$). This promises that the D$_1$ molasses scheme may be exploited to image $^6$Li atoms in optical lattices with single-site resolution \cite{greiner, bloch_im}. At the end of our typical experimental runs we can produce either pure molecular Bose-Einstein condensates (mBEC) of 5$\times10^5$ molecules and binary mixtures of degenerate Fermi gases of 3.5$\times10^{5}$ atoms per spin state at $T/T_{F}<0.1$, where T$_F$ is the Fermi temperature. We believe that this scheme is a convenient method of producing large and highly degenerate fermionic clouds of $^6$Li without the need of any additional coolant atomic species \cite{fifty_fold}.\\
Our experimental sequence starts with loading a $^6$Li MOT operating on D$_2$ (2S$_{1/2}\rightarrow$2P$_{3/2}$) optical transitions. We load about 2$\times$10$^9$ atoms via standard laser cooling techniques. The MOT light configuration consists of three mutually orthogonal retro-reflected laser beams with $1/e^{2}$ radius of about 1.5~cm, and  peak intensity of about 7 I$_S$, where I$_S$=2.54 mW/cm$^2$ is the saturation intensity of the D$_2$ transition. Each beam contains both cooling (-9$\Gamma$ detuned from the F=3/2$\,\rightarrow\,$F'=5/2 transition) and repumper (-6$\Gamma$ detuned from the F=1/2$\,\rightarrow\,$F'=3/2 transition, where $\Gamma=2\pi\times5.87$~MHz) light. The power of the cooling light relative to the repumper one is 3:2. 
The large detuning is chosen to maximize the number of trapped atoms, limiting the initial temperature of the MOT to about 2.5~mK. We cool and compress the D$_2$ MOT by reducing the intensity of both the repumper and cooling light to about 1$\%$ of the initial value, while simultaneously decreasing the detuning of both to -3$\Gamma$. Here, the temperature of the cloud drops to about T$_0$=500~$\mu$K in 2 ms and N$_{0}$\,=$\,1.6\times10^9$ atoms remain. At this point we turn off the D$_2$ MOT lights and the magnetic quadrupole field while turning on the D$_1$ molasses. The molasses is comprised of cooling (F=3/2$\,\rightarrow\,$F'=3/2) and repumper lights (F=1/2$\,\rightarrow\,$F'=3/2), both blue detuned with respect to the resonances (see Fig.~\ref{fig:schemalivelli}). The D$_2$ and the D$_1$ lights are provided by two different laser sources, independently controlled by two acousto-optic modulators (AOMs) which act as fast switches. The two lasers inject into the same tapered amplifiers, so the same optical components are used for realizing D$_2$ MOT and D$_1$ molasses, and no further alignment is needed.\\
We have tested the performance of D$_1$ cooling by varying the molasses parameters. The temperature and the number of atoms are determined after time-of-flight expansion, via absorption imaging resonant with the D$_2$ transition. 
In Fig.~\ref{fig:Fano}~(a) we show the evolution of temperature and atom number after 2~ms of gray molasses as a function of the relative detuning, $\delta=\delta_{1}-\delta_{2}$. Here, $\delta_{1}$ and $\delta_{2}$ are the detuning of the D$_1$ repumper and cooling lasers from the F=1/2$\,\rightarrow\,$F'=3/2 and F=3/2$\,\rightarrow\,$F'=3/2 transitions, respectively (see Fig.~\ref{fig:schemalivelli}). Since the effectiveness of the D$_1$ cooling strongly depends on the ratio of the repumper and cooling intensities \cite{PhysRevA.87.063411,Chevy-EPL-12,PhysRevA.88.053407,Bourdel-EPLdraft-13} in this paper we fix $I_{rep}\simeq 0.2I_{cool}$, the experimentally determined value that gives the maximum cooling efficiency.
Under this condition, the dependence of temperature on $\delta$ exhibits an asymmetric Fano profile with sub-natural width in a narrow range around the Raman resonance ($\delta$=0). This is an evident signature of the emergence of a quantum interference effect \cite{fano}. Indeed as the laser fields match the Raman condition, the temperature drops to its minimum value, T=40.5(1.0)~$\mu$K, with a cooled fraction N/N$_0$ of 75~$\%$, as a consequence of both the Sisyphus effect on the blue of the $F\rightarrow F'=F$ transition \cite{PhysRevA.87.063411} and the formation of a coherent dark state \cite{PhysRevA.87.063411,Chevy-EPL-12,PhysRevA.88.053407,Bourdel-EPLdraft-13}. For $\delta$ slightly blue-detuned from the resonance we observe instead a strong heating accompanied by atom loss, as discussed in \cite{PhysRevA.87.063411,Chevy-EPL-12,PhysRevA.88.053407,Bourdel-EPLdraft-13}. Away from the resonance the temperature and the number of atoms reach stationary values due to the Sisyphus effect alone \cite{PhysRevA.87.063411}. Remarkably, we observe that the efficiency of the gray-molasses depends on the temperature T$_0$ of the cloud before the molasses is applied. In Fig.~\ref{fig:Fano} (b) we show the behavior of the cooled fraction of atoms N/N$_0$ versus T$_0$, for $\delta$=0. The cooled fraction reaches 100$\%$ for initial temperatures below 150~$\mu$K, close to the Doppler limit, while the final temperature does not depend on T$_0$. Interestingly, at the Raman condition, the effect of the molasses is almost insensitive to $\delta_{2}$ in a broad range of detuning, from 4 to 8 $\Gamma$, as shown in Fig.~\ref{fig:Fano} (c).\\
At the end of the gray-molasses stage, about 85$\%$ of the atoms are in the $\left|F=1/2\right\rangle$ manifold, a value determined by I$_{rep}$/I$_{cool}$. In our optimized cooling conditions, we estimate a peak phase-space density of 2$\times$10$^{-5}$, about 50 times larger than that obtained with only the D$_2$ cooling stages. Such a high phase-space density is desirable when transferring the atoms into an optical potential.\\
Our single-beam optical dipole trap (ODT) is generated by a 200~W  multi-mode ytterbium fiber laser with a central wavelength  of 1073~nm. Its initial power is set to 120~W. The laser is focused on the atoms with a waist of 42 $\mu$m, at an angle of about 15$^\circ$ with respect to one of the horizontal MOT beams.
To increase the trapping volume, we create a time-averaged optical potential by modulating the frequency and amplitude of the ODT's control AOM at a frequency much greater than the natural trapping frequency \cite{timeaveraged3}. This results in an elliptic Gaussian-shaped beam with waists of about 42~$\mu$m (along gravity) $\times$ 85$\mu$m. The estimated initial trap depth is on the order of 1~mK. 
The optical potential is ramped up over 5~ms during the D$_2$ cooling stage and it is fully on by the time the D$_1$ phase is applied. Mode-matching between the MOT and the ODT is optimized by unbalancing the relative intensity of MOT beams, creating an oblate cloud perpendicular to gravity and elongated in the direction of propagation of the ODT beam. Despite this strong intensity anisotropy, the performance of the molasses is almost the same as that in the balanced configuration.\\
\begin{figure}[t]
\includegraphics[width=8 cm,clip]{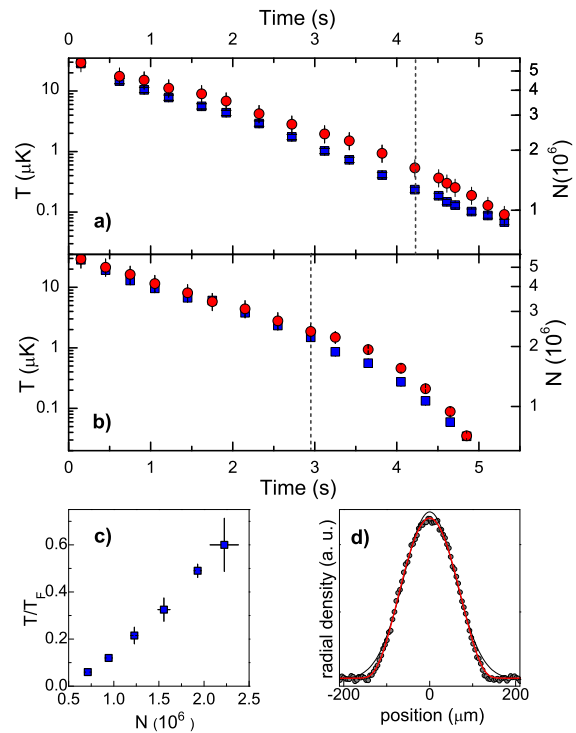}
\caption{(Color online) (a) Total atom number N (blue squares) and temperature T (red circles) of the atoms during forced evaporation in the optical trap at B=800~G. The dotted line marks the evaporation time for which T/T$_c$=1. (b) Total atom number N (blue squares) and temperature T (red circles) of the atoms during forced evaporation in the optical trap at B=300~G. The dotted line indicates the evaporation time for which T/T$_F$=1. Above this value, the temperature is measured after expansion close to the zero-crossing, while T$_F$ is determined by the mean value of N and the measured trap frequencies. For T/T$_F$$<$1 a surface fit to a polylog function is used to determine T/T$_F$. (c) T/T$_F$ versus N approaching the end of the evaporation ramp, where T/T$_F$=0.06(1). (d) Comparison between a Gaussian (black line) and the finite-temperature Fermi distribution fit (red line) for a typical experimental profile corresponding to T/T$_F$=0.06(1). In (a), (b) and (c) error bars account for both statistical (five independent measurements) and 10$\%$ systematic uncertainties.
}
\label{fig:EvapFermi}       
\end{figure}
To test the feasibility of D$_1$ cooling in the presence of the strong light field of the optical potential, we first measure the light shifts of the D$_1$ transitions (cooling and repumper) as a function of the ODT power. We obtain from a combined fit a slope of +8.2(7)~MHz/(MW/cm$^2$), corresponding to a shift of about 16~MHz ($\sim$3$\Gamma$) for our initial trapping intensity. The uncertainty is mostly due to the systematic uncertainty (10~$\%$) in the estimation of the trap intensity. This measurement indicates that D$_1$ molasses can properly work in the ODT, provided that the absolute detuning accounts for these light shifts, remaining in the range shown in Fig.~\ref{fig:Fano}~(c).\\ We find the largest number of atoms in the optical trap when cooling for 2~ms on the D$_1$ transition with $\delta$=-0.2$\Gamma$ (slightly off the Raman condition) and $\delta_2$=5.4$\Gamma$. These parameters correspond to the maximum atom number captured by the gray-molasses (see  Fig.~\ref{fig:Fano} (a)), and they result in temperatures  well below the initial trap depth. The total number of atoms transferred into the optical trap is typically N=2$\times$10$^7$ at T=135(5)~$\mu$K. To test the efficiency of the gray-molasses on atoms in the optical trap, we repeat the measurement of Fig.~\ref{fig:Fano}(a), after 25~ms following the end of the ODT loading. In particular, we apply a second D$_1$ cooling stage lasting 300~$\mu$s and we vary the relative detuning $\delta$, with $\delta_2$=5.4 $\Gamma$. The results are shown in Fig.~\ref{fig:Fano}(d). Qualitatively, the behavior of temperature and atom number following cooling is similar to that measured in the absence of the optical trap. In particular, after the second D$_1$ stage, the temperature drops to a minimum value, in this case T=80(5)~$\mu$K, for $\delta$=0. This indicates that D$_1$ cooling is still efficient even in the presence of the ODT's high-intensity laser field. The minimum temperature achieved in the ODT is almost a factor of two higher than that measured without optical confinement, and it is accompanied by a broadening of the Fano profile around $\delta$=0. We ascribe this behavior to the large atom density inside the optical trap, which may limit the efficiency of D$_1$ cooling \cite{nota2}. In the optical trap, the minimum temperature corresponds to the maximum cooled fraction (100~$\%$). This because the initial temperature of the atoms collected in the ODT is about 135~$\mu$K, sufficiently low to allow an effective cooling of  all the atoms (see Fig.~\ref{fig:Fano}(b)). As a result, our optimal experimental procedure to load the atoms into the optical trap consists of two different stages of D$_1$ gray-molasses cooling. The first one lasting 2~ms at $\delta$=-0.2$\Gamma$, maximizes the number of trapped atoms, while the second of 300~$\mu$s at $\delta$=0 cools the sample to T=80(5)~$\mu$K. This second stage is followed by a 25~$\mu s$ hyperfine pumping to the $\left|F=1/2\right\rangle$ manifold, achieved by turning off the D$_1$ repumper light before the cooling light. This hyperfine pumping stage increases the temperature by about 10$\%$. After pumping, we ramp the Feshbach field in about 30~ms up to 840~G, close to the center of the Feshbach resonance \cite{jochim}. To limit thermal lensing effects that result from the high laser intensity, we rapidly start the evaporation by reducing the laser power to 30~W in 500~ms. Multiple radio-frequency sweeps resonant with the $\left|F=1/2,m_F=\pm1/2\right\rangle$ transition create the favorable incoherent balanced spin mixture of these two Zeeman levels. In what follows, we denote these states as $\left|1\right\rangle$ and $\left|2\right\rangle$. After the first evaporation ramp we typically have 1$\times$10$^7$ atoms per spin state at T$\simeq$~30$~\mu$K.\\ 
To produce a molecular BEC, we perform evaporation of the $\left|1\right\rangle$-$\left|2\right\rangle$ mixture at 800 G, where the s-wave scattering length $\it{a_{12}}$ is on the order of 11000~$\it{a}_0$ \cite{jochim}, where $\it{a}_0$ is the Bohr radius. Molecules are formed via three-body recombination processes as soon as the temperature of the cloud becomes comparable with the molecular binding energy \cite{Grimm}. At T=T$_c$=210(20)~nK, we observe the onset of condensation for N$_{mol}\simeq1\times10^6$ molecules. To resolve the condensate fraction, we reduce the inter-particle interaction by adiabatically sweeping the magnetic field to 690~G, where $\it{a_{12}}\simeq$1400~$\it{a}_0$ \cite{jochim}. The molecules are then released from the trap and imaged at this magnetic field using a closed optical transition. 
At T$_c$, the measured trap frequencies are $\omega_x=2\pi\times$8.2(1)~Hz, $\omega_y=2\pi\times$111(3)~Hz, and $\omega_z=2\pi\times$239(2)~Hz, where the lowest frequency is given by the magnetic curvature of our Feshbach coils. By reducing the trap depth further, we observe the formation of a mBEC of N$_{mol}\simeq5\times10^5$ molecules with no discernible thermal component. A similar scheme is exploited to create a unitary Fermi gas at the center of the resonance. We observe ultracold clouds of about 2$\times10^6$ particles at a temperature corresponding to T$_c$, when sweeping to the molecular side of the resonance.\\ 
The strategy to create weakly-interacting Fermi gases is slightly different: we evaporate the $\left|1\right\rangle$-$\left|3\right\rangle$ spin-mixture, where the $\left|3\right\rangle$ corresponds to the $\left|F=3/2,m_F=-3/2\right\rangle$ level at low magnetic fields. The mixture is created at T$\simeq$30~$\mu$K by transferring  100$\%$ of the atoms from the state $\left|2\right\rangle$ to the $\left|3\right\rangle$ by a radio-frequency sweep. The mixture is then evaporated at 300~G, where the $\left|1\right\rangle$-$\left|3\right\rangle$ s-wave scattering length is $\it{a}_{13}\simeq-\mbox{880}\it{a}_0$, almost three times larger than $\it{a}_{12}\simeq-\mbox{290}\it{a}_0$ \cite{jochim}, strongly enhancing the efficiency of evaporation. The axial confinement at this magnetic field is provided by a magnetic curvature generated by an additional pair of coils. 
In Fig.~\ref{fig:EvapFermi} (a) and (b), we compare evaporation trajectories for the $\left|1\right\rangle$-$\left|2\right\rangle$ mixture at 800~G (a) and the $\left|1\right\rangle$-$\left|3\right\rangle$ mixture at 300~G (b). The two trajectories are similar, demonstrating the efficient thermalization between the $\left|1\right\rangle$-$\left|3\right\rangle$ states. After 3 seconds of forced evaporation, the system enters the degenerate regime with N$_{\left|1\right\rangle}$=N$_{\left|3\right\rangle}$=2$\times$10$^6$ atoms. After a further 2 seconds, we produce a highly degenerate Fermi gas of N$_{\left|1\right\rangle}$=N$_{\left|3\right\rangle}$=3.5$\times$10$^5$ atoms at T/T$_F\simeq 0.06(1)$, where T$_F$ is the Fermi temperature defined as $k_BT_F=\hbar\varpi(6N_{i})^{1/3}$, where 
$\varpi=(\omega_x\omega_y\omega_z)^{1/3}$ (see Fig.~\ref{fig:EvapFermi} (c)). Here the measured trapping frequencies are $\omega_x$=2$\pi\times$12.4(1)~Hz, $\omega_y$=2$\pi\times$111.5(2)~Hz, $\omega_z$=2$\pi\times$231(3)~Hz. The degree of degeneracy is extracted by fitting the density profiles of the atomic samples with a bidimensional finite-temperature Fermi distribution \cite{varenna} (Fig.~\ref{fig:EvapFermi}(d)).\\
In conclusion, we have demonstrated an all-optical scheme to produce large and deeply degenerate $^6$Li gases. Our method is based on the combination of D$_1$ gray molasses and optical trapping. This sub-Doppler cooling mechanism allows us to lower the initial MOT temperature to about 40~$\mu$K without significant atom loss, obtaining ideal conditions for loading the atoms into deep optical potentials. We demonstrate that this gray-molasses scheme is robust and that it works efficiently in the presence of such intense infrared trapping laser fields.
Thanks to these ingredients, we have produced pure Bose-Einstein condensates of up to 5$\times$10$^5$ molecules and degenerate Fermi gases of about 10$^6$ atoms below T/T$_F<$ 0.1 with a typical duty cycle of 11 seconds. These numbers can be increased further by engineering larger volume optical potentials, such as optical resonators \cite{Grimm}. In the future we will investigate in further detail the possibility of using the D$_1$ molasses as a tool to image $^6$Li atoms in deep optical potentials. We believe that our results will be important for experiments aimed at implementing quantum Hamiltonians in optical lattices with ultracold atomic fermions \cite{esslinger,bloch_2}.

We thank A. Morales and A. Trenkwalder for useful discussions and contributions to the experiment, and R. Ballerini and A. Hajeb and the members of the LENS electronic workshop for technical support during the building stage of the experiment. Special acknowledgments to the LENS Quantum Gases  group. We also are grateful to F. Chevy for useful insights on D$_1$ molasses.  E. P. has been supported by MIT's IROP program. This work was supported under the ERC Grant No.307032 QuFerm2D.


\begin{thebibliography}{30}
\bibitem{nota2} We have checked that applying D1 cooling in much lower density clouds results in colder samples down to 50~$\mu$K.
\end{thebibliography}
\end{document}